\documentclass{article}
\usepackage{graphicx} 
\usepackage[utf8]{inputenc}
\usepackage{hyperref}
\usepackage{amsmath, amssymb}

\usepackage[backend=biber, style=chicago-authordate, natbib=true]{biblatex}
\title{Crypto Market Analysis \& Ethereum Blockchain Real-Estate Business Protocol}
\author{Sid Bhatia}
\date{April 2024}

\usepackage[
    top=2cm,
    bottom=2cm,
    left=2cm,
    right=2cm,
    headheight=17pt, 
    includehead,
    includefoot,
    heightrounded, 
]{geometry}

\usepackage[backend=biber, style=chicago-authordate, natbib=true]{biblatex}
\begin{document}

\begin{titlepage}
    \centering
    Stevens Institute of Technology \\
    \vspace{1.5cm}
    \vspace{4cm}
    {\huge\bfseries Crypto Market Analysis \& Real-Estate Business Protocol Proposal \\}
    \vspace{0.5cm}
    {\Large Application of Ethereum Blockchain \\}
    \vspace{0.5cm}
    \vfill
    \textsc{\Large Sid Bhatia, Samuel Gedal, Himaya Jeyakumar \\ Grace Lee, Ravinder Chopra, Daniel Roman, Shrijani Chakroborty \\}
    \vfill
    {\large 26 April 2024}
\end{titlepage}

\newgeometry{
    margin=0.8in
}

\newpage
\tableofcontents
\newpage

\section{Introduction}

In the dynamic realm of financial technology, blockchain and cryptocurrencies represent two of the most significant innovations that have reshaped how transactions are conducted and assets are managed globally (\cite{futurecryptocurrencies2022}). This paper delves into a dual-focused analysis and proposal. Firstly, we conduct a thorough market analysis of a select group of cryptocurrencies, each chosen for its unique role and impact within the broader digital currency landscape. The cryptocurrencies under review include Bitcoin, often regarded as the progenitor of all digital currencies; Ethereum, notable for its robust smart contract capabilities; XRP, designed primarily for rapid financial transactions; Dogecoin, which began as a meme but has since gained substantial practical application; and Tether, a stablecoin tied to the US dollar, offering a less volatile refuge within the highly fluctuant crypto market (\cite{evolutioncryptomarket2017}).

This study not only examines the price trends, volatilities, and inter-cryptocurrency correlations but also assesses the impact of significant market events, such as the FTX bankruptcy, on these digital assets (\cite{ftxresponse2023}). The insights garnered from this analysis aim to provide a granular understanding of how various cryptocurrencies react to internal and external pressures, influencing investor sentiment and market dynamics.

Following the market analysis, the second focus of this paper introduces an innovative business proposal leveraging blockchain technology. This proposal outlines a new protocol for real estate transactions, allowing property deeds to be securely managed and transferred without the need for traditional intermediaries such as lawyers and brokers. By employing blockchain technology, this protocol seeks to revolutionize the real estate market by enhancing transparency, reducing transaction costs, and simplifying the transaction process for buyers and sellers across the globe (\cite{blockchainrealestate2021}).

Through comprehensive analysis and forward-thinking proposals, this paper contributes to the ongoing discussions surrounding the application of blockchain technology in traditional sectors, proposing not only a new way to understand cryptocurrencies in relation to the traditional financial markets but also offering a practical application that addresses real-world challenges in real estate transactions.

\hrulefill

\section{Part I: Crypto Market Analysis}

\subsection{Overview}

This analysis encompasses a selection of five distinct cryptocurrencies, each representing a unique facet of the current digital currency ecosystem. Our selected cryptocurrencies include: \textbf{Bitcoin (BTC)}, recognized as the original and most well-known cryptocurrency; \textbf{Ethereum (ETH)}, noted for its advanced smart contract capabilities; \textbf{XRP}, developed by Ripple Labs with a focus on rapid digital payments; \textbf{Dogecoin (DOGE)}, which has evolved from a meme into a cryptocurrency with practical uses in tipping and donations; and \textbf{Tether (USDT)}, a stablecoin that introduces a measure of stability in the otherwise volatile cryptocurrency market. This diverse selection aims to cover a broad spectrum of functionalities, market positions, and technological innovations within the crypto space, providing a comprehensive overview of its varied applications and implications.

\subsection{Detailed Overview and Crypto Protocol}

\subsubsection{Bitcoin (BTC)}

\textbf{Overview:} Introduced in 2009 by an entity under the pseudonym Satoshi Nakamoto, Bitcoin stands as the inaugural cryptocurrency, designed to operate as a decentralized digital currency without the oversight of a central authority. Transactions are conducted directly between users through the peer-to-peer Bitcoin network.

\textbf{Protocol:} Bitcoin's network is underpinned by a proof-of-work (PoW) protocol, wherein miners employ significant computational resources to solve intricate mathematical problems, thus validating transactions and securing the network, with new bitcoins awarded as a mining reward.

\textit{For more details see \cite{nakamoto2009bitcoin}.}

\subsubsection{Ethereum (ETH)}

\textbf{Overview:} Launched in 2015, Ethereum transcends the conventional definition of a cryptocurrency. It serves as a platform for the development of decentralized applications (DApps) through smart contracts, aiming to democratize access to a decentralized financial system.

\textbf{Protocol:} Initially based on a proof-of-work mechanism similar to that of Bitcoin, Ethereum is transitioning to a proof-of-stake (PoS) model with its Ethereum 2.0 update, which promises enhanced scalability and reduced energy consumption.

\textit{Refer to \cite{buterin2015ethereum} for additional insights.}

\subsubsection{XRP (Ripple)}

\textbf{Overview:} Created by Ripple Labs in 2012, XRP is central to a digital payment protocol that surpasses its identity as a mere cryptocurrency. It facilitates rapid payment settlements across the network.

\textbf{Protocol:} The XRP Ledger utilizes a consensus protocol that does not rely on the traditional blockchain mining process; instead, it achieves consensus through a network of independent validating servers that constantly compare transaction records.

\textit{See \cite{ripple2012xrp} for further information.}

\subsubsection{Dogecoin (DOGE)}

\textbf{Overview:} Originating as a humorous take on the cryptocurrency phenomenon in 2013, Dogecoin was inspired by the "Doge" meme featuring a Shiba Inu. It has since cultivated a community focused on using the cryptocurrency for charitable contributions and tipping online content creators.

\textbf{Protocol:} Dogecoin operates on a less energy-intensive proof-of-work algorithm derived from Litecoin, known as Scrypt, facilitating faster transaction processing.

\textit{Detailed information available at \cite{dogecoin2013meme}.}

\subsubsection{Tether (USDT)}

\textbf{Overview:} Introduced in 2014, Tether represents a stablecoin that is tethered to the US dollar, aiming to meld the flexibility of cryptocurrencies with the stability of fiat currency.

\textbf{Protocol:} Tether supports a hybrid use of protocols, operating on the Omni Layer of the Bitcoin blockchain and as an ERC-20 token on the Ethereum blockchain, among other blockchain platforms.

\textit{Further details can be found in \cite{tether2014usdt}.}

\

These cryptocurrencies were chosen to provide a diverse perspective on the various applications, market usage, and technological advancements within the broader cryptocurrency environment. From January 1, 2022, to December 31, 2022, our study observed no missing data, ensuring the completeness and reliability of the analysis conducted during this period.

\subsection{Crypto Selection Rationale}

The selection of cryptocurrencies for this study was informed by a multifaceted rationale emphasizing diversity, technological innovation, community engagement, and market stability. Each cryptocurrency was chosen not only for its unique position within the market but also for its contribution to advancing the blockchain technology landscape.

\textbf{Diversity and Relevance:} Bitcoin and Ethereum are selected as foundational pillars within the cryptocurrency domain, illustrating the broad spectrum of blockchain applications. Bitcoin, often hailed as the original cryptocurrency, has pioneered the concept of a decentralized digital currency and enjoys widespread adoption and recognition. Ethereum, on the other hand, extends the utility of blockchain beyond mere financial transactions through its support for smart contracts, thereby catalyzing a plethora of decentralized applications (DApps). This diversity underscores the significant role these currencies play in the ongoing development and maturation of the cryptocurrency market.

\textbf{Technological Diversity:} XRP and Tether were chosen to highlight the technological diversity within blockchain implementations. XRP, developed by Ripple, is notable for its rapid transaction capabilities and minimal energy consumption, diverging from the traditional mining-based consensus used by currencies like Bitcoin. Similarly, Tether introduces a model of stability in the highly volatile cryptocurrency market by being pegged to the US dollar, showcasing a unique application of blockchain technology in creating stablecoins that mitigate the price volatility typically associated with cryptocurrencies.

\textbf{Community and Innovation:} Dogecoin exemplifies the impact of community on the value and adoption of a cryptocurrency. Originating as a meme, Dogecoin has transcended its initial novelty to foster a robust community that actively engages in tipping and charitable activities through the currency. This aspect highlights the role of societal and cultural dynamics in shaping the cryptocurrency landscape, emphasizing the importance of community-driven development and innovation.

\textbf{Market Stability and Innovations:} Finally, the inclusion of Tether also addresses the critical challenge of market stability. By anchoring its value to a stable fiat currency, Tether offers a pragmatic solution to the issue of volatility, which is a pervasive concern for investors in cryptocurrencies like Bitcoin and Ethereum. This approach not only facilitates greater market stability but also enhances the practicality of cryptocurrencies for everyday transactions and financial applications.

\

Collectively, these selections provide a comprehensive overview of the current state and potential future directions of blockchain technology, illustrating a spectrum of use cases from foundational cryptocurrencies to innovative adaptations addressing specific market needs.

\subsection{Market Analysis}

\begin{figure}[ht]
\centering
\includegraphics[width=\linewidth]{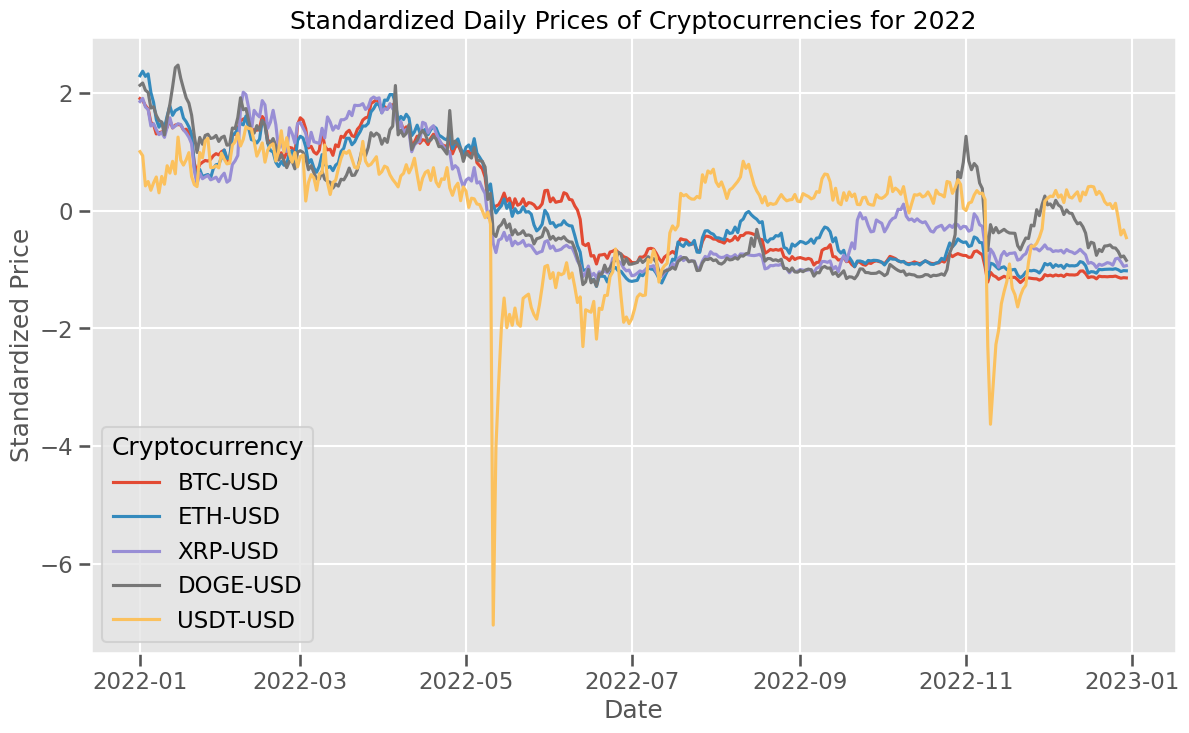}
\caption{Standardized Daily Prices of Cryptocurrencies for 2022}
\label{fig:std-daily-prices}
\end{figure}

\subsubsection{Price Trend Analysis}

The analysis of standardized price trends of Bitcoin (BTC), Ethereum (ETH), XRP, Dogecoin (DOGE), and Tether (USDT) throughout 2022 reveals several key insights into the dynamics of the cryptocurrency market:

\paragraph{Correlated Movements:} The data illustrates that most cryptocurrencies exhibited closely correlated movements over the course of the year. Such correlation is indicative of the substantial influence exerted by broader market forces and global economic events on the cryptocurrency market as a whole, driving collective swings in investor sentiment—whether bullish or bearish.

\paragraph{Volatility Across Assets:} The degree of volatility varied significantly among the analyzed cryptocurrencies. Bitcoin and Ethereum experienced relatively moderate fluctuations, maintaining tighter price bands, while Dogecoin displayed higher volatility, characterized by more pronounced peaks and troughs. This disparity in volatility underscores the differential market perceptions and investor bases of these assets.

\paragraph{Stablecoin Anomaly:} An unexpected anomaly was observed in the price trend of Tether (USDT), particularly in May 2022, where it deviated markedly from its expected stable trajectory. Such a divergence, given the design of stablecoins to maintain parity with a peg (e.g., USD), suggests potential extraordinary events, data reporting inaccuracies, or underlying issues with the stability mechanisms during that period.

\paragraph{Market Recovery Ability:} Following significant market dips, the cryptocurrencies demonstrated varying degrees of recovery. Bitcoin and Ethereum showed robust resilience and recovery capabilities compared to Dogecoin. This variation could reflect differing levels of market confidence and inherent stability within these digital assets.

\paragraph{Stablecoin's Peculiar Trend:} Assuming the accuracy of the observed sharp decline in USDT's value, this could represent a period of intense market stress or a temporary disruption in the stablecoin’s dollar peg. However, such incidents are generally ephemeral, as corrective mechanisms typically restore stability swiftly, aligning with the observed rapid return to normalcy.

\

From the analysis of these price trends, it is evident that while cryptocurrencies are interconnected and respond collectively to market shifts, individual assets exhibit distinct behaviors influenced by their specific market dynamics, investor sentiment, and technological foundations. The peculiar movement observed in Tether’s price trend during the analyzed period merits further investigation to ascertain the causes and implications of such an anomaly.

\subsubsection{Volatility Analysis}

The study of volatility in cryptocurrency markets provides crucial insights into the risks and stability of digital assets. By calculating daily returns and examining their standard deviations, we can gauge the unpredictability associated with each cryptocurrency and identify the factors contributing to these dynamics.

\paragraph{Dogecoin (DOGE-USD):} Dogecoin exhibits the highest volatility among the cryptocurrencies analyzed, with a standard deviation of approximately \textbf{5.64\%}. This elevated volatility can primarily be attributed to its relatively low price per unit, which renders it more susceptible to significant percentage changes on a per-unit basis. Moreover, Dogecoin's price is notably influenced by social media trends and possesses comparatively less market liquidity than more established cryptocurrencies. These elements combine to increase its price volatility, reflecting the substantial impact of retail investor sentiment and speculative trading on its market behavior.

\paragraph{Tether (USDT-USD):} In stark contrast, Tether shows the lowest volatility, with a standard deviation near \textbf{0.03\%}. As a stablecoin, Tether is explicitly designed to be pegged to a fiat currency, specifically the US dollar, and maintains a stable value through various regulatory and technological mechanisms. This stability is critical for its role in providing a safe haven during market turbulence and for facilitating transactions where volatility can be a deterrent.

\paragraph{Bitcoin (BTC-USD) and Ethereum (ETH-USD):} Both Bitcoin and Ethereum exhibit moderate levels of volatility, reflecting their established presence in the market and larger capitalizations. These factors typically confer higher liquidity and result in less drastic percentage changes in daily prices.

\paragraph{Benchmark Volatility Analysis:}
Comparing the volatility of cryptocurrencies with traditional financial markets, such as the S\&P 500, highlights the unique risk profiles inherent to digital assets. The S\&P 500, with a volatility of \textbf{1.00\%}, offers a contrast to the higher volatility levels seen in cryptocurrencies, underscoring the potential for greater price stability in traditional equity markets.

\paragraph{Market Implications:} This variability in volatility, especially when benchmarked against traditional indices like the S\&P 500, illustrates the diverse nature of cryptocurrency markets. While stablecoins like Tether aim to minimize price fluctuations, other cryptocurrencies such as Dogecoin and Bitcoin exhibit a range of volatilities, heavily influenced by investor sentiment, liquidity, and their roles within the digital economy. 

The higher volatility of cryptocurrencies compared to traditional markets like the S\&P 500 underscores their speculative nature and the heightened risks they pose, which investors must navigate carefully. This analysis emphasizes the importance of strategic risk assessment and portfolio diversification to mitigate the inherent volatility of cryptocurrencies.

\subsubsection{Correlation Analysis}

\begin{figure}[ht]
\centering
\includegraphics[width=\linewidth]{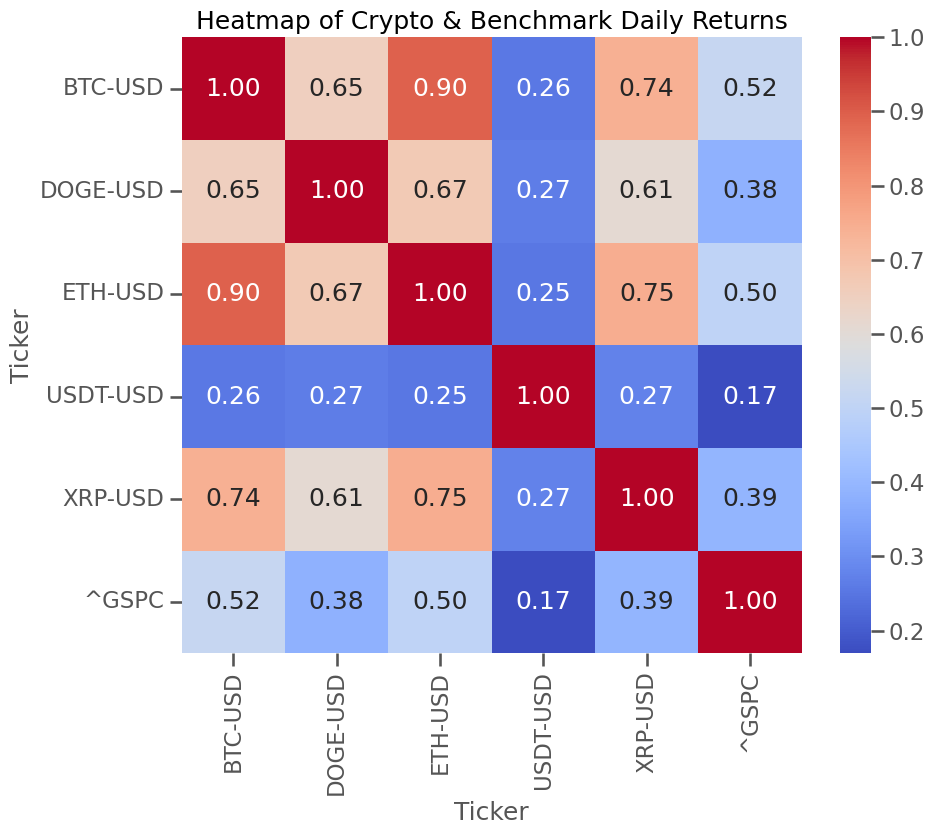}
\caption{Correlation Matrix Heatmap of Cryptocurrencies for 2022}
\label{fig:corr-matrix-heatmap}
\end{figure}

A comprehensive examination of the correlation matrix for daily returns of Bitcoin (\textbf{BTC}), Ethereum (\textbf{ETH}), XRP, Dogecoin (\textbf{DOGE}), and Tether (\textbf{USDT}) in tandem with the S\&P benchmark (\textbf{GSPC}) elucidates the interrelationships among these prominent cryptocurrencies:

\paragraph{High Correlations:}
\begin{itemize}
    \item \textbf{Bitcoin and Ethereum:} Exhibiting a correlation coefficient of \textbf{0.90}, BTC and ETH demonstrate a very strong positive correlation, indicating that these cryptocurrencies often move in tandem. This strong linkage is primarily due to their predominant positions in the market, where both are frequently influenced by similar economic factors, investor sentiments, and regulatory developments.
    \item \textbf{Ethereum and XRP:} With a correlation of \textbf{0.75}, movements in Ethereum frequently correlate closely with those in XRP, suggesting overlapping functionalities and investor bases that react similarly to market stimuli in these two platforms.
\end{itemize}

\paragraph{Moderate Correlations:}
\begin{itemize}
    \item \textbf{XRP with Bitcoin and Dogecoin:} XRP displays moderate correlations of \textbf{0.74} with BTC and \textbf{0.61} with DOGE. These correlations suggest a level of synchronicity, albeit influenced by distinct market dynamics and external factors specific to each cryptocurrency.
    \item \textbf{Dogecoin with Ethereum and Bitcoin:} Correlation coefficients of \textbf{0.67} with ETH and \textbf{0.65} with BTC for Dogecoin indicate a moderate degree of correlation, influenced by broader market trends that impact all cryptocurrencies, though each responds according to its unique market niche and investor behavior.
\end{itemize}

\paragraph{Lower Correlations with Tether:}
\begin{itemize}
    \item \textbf{All Cryptocurrencies with Tether:} Tether, being a stablecoin tied closely to the US dollar, shows significantly lower correlation coefficients with BTC (\textbf{0.26}), ETH (\textbf{0.25}), XRP (\textbf{0.28}), and DOGE (\textbf{0.27}). This fundamental difference in design and purpose—aimed at providing stability—results in less synchronized movements with the more speculative cryptocurrency assets.
\end{itemize}

\paragraph{Benchmark Correlation Comparison:} Comparison with traditional financial markets, specifically through a correlation study with the S\&P 500, reveals additional insights. While cryptocurrencies such as BTC and ETH show the highest correlation with each other, they exhibit only moderate correlation levels with the S\&P 500, with BTC showing the highest correlation at \textbf{0.52}. This suggests that while cryptocurrencies do move somewhat in sync with traditional financial markets, they retain distinct market dynamics that set them apart.

\paragraph{Market Implications:} These findings highlight the diverse correlation landscapes within the cryptocurrency markets, where strong intra-crypto correlations contrast with more moderate interactions with traditional financial indices. This divergence underscores the necessity for investors to consider the unique correlation patterns when diversifying portfolios or implementing hedging strategies. The mixed correlation profiles suggest both opportunities and risks, as cryptocurrencies can offer portfolio diversification benefits due to their partial independence from traditional market movements.

\subsection{FTX Delta Analysis}

\subsubsection{Event Overview}
In mid-November 2022, the cryptocurrency exchange FTX filed for bankruptcy, triggering significant disturbances across the cryptocurrency markets. This event was exacerbated by the resignation of its CEO, Sam Bankman-Fried, further destabilizing the market's confidence.

\subsubsection{FTX Impact on 11/11/22}
\begin{table}[h]
\centering
\begin{tabular}{|l|c|}
\hline
\textbf{Ticker} & \textbf{Impact on Nov 11} \\
\hline
BTC-USD  & -3.14\% \\
DOGE-USD & -5.46\% \\
ETH-USD  & -0.94\% \\
USDT-USD & 0.04\%  \\
XRP-USD  & -2.92\% \\
GSPC   & 1.00\%  \\
\hline
\end{tabular}
\caption{FTX Impact on Cryptocurrency Prices on November 11, 2022}
\label{tab:ftx-impact-nov11}
\end{table}

\subsubsection{Immediate Impact}
The immediate repercussions of the bankruptcy announcement on November 11, 2022, were starkly evident across various cryptocurrencies:
\begin{itemize}
    \item \textbf{Bitcoin (BTC)} and \textbf{Ripple (XRP)} each faced notable declines, with Bitcoin falling by \textbf{-3.14\%} and XRP by \textbf{-2.92\%}.
    \item \textbf{Ethereum (ETH)} exhibited relative resilience, with a modest decline of \textbf{-0.94\%}, reflecting its robust market presence and investor confidence.
    \item \textbf{Dogecoin (DOGE)} experienced the most significant drop of \textbf{-5.46\%}, illustrating its susceptibility to market shocks.
    \item \textbf{Tether (USDT)}, maintaining its stability, changed insignificantly by \textbf{+0.04\%}, underscoring its role as a stabilizing force within the volatile cryptocurrency environment.
\end{itemize}

\subsubsection{FTX Impact in November 2022}
\begin{table}[h]
\centering
\begin{tabular}{|l|c|}
\hline
\textbf{Ticker} & \textbf{Change in Nov 2022} \\
\hline
BTC-USD  & -16.19\% \\
DOGE-USD & -25.05\% \\
ETH-USD  & -17.98\% \\
USDT-USD & 0.01\%   \\
XRP-USD  & -12.02\% \\
GSPC   &  5.81\%   \\
\hline
\end{tabular}
\caption{Monthly Impact of FTX Bankruptcy on Cryptocurrency Prices in November 2022}
\label{tab:ftx-impact-nov}
\end{table}

\subsubsection{Long-Term Impact}
The extended impact throughout November painted a grim picture of recovery challenges:
\begin{itemize}
    \item Major cryptocurrencies like \textbf{BTC}, \textbf{ETH}, and \textbf{XRP} recorded substantial declines of \textbf{-16.19\%}, \textbf{-17.98\%}, and \textbf{-12.02\%} respectively.
    \item \textbf{DOGE} was particularly hard hit, plummeting by \textbf{-25.05\%}, marking the highest vulnerability among the group.
    \item Conversely, \textbf{USDT} showed remarkable stability with only a \textbf{0.01\%} change, reinforcing its value proposition as a hedge against volatility.
\end{itemize}

\subsubsection{Benchmark and Market Performance Comparison}
The correlation and impact studies reveal that while the cryptocurrency market suffered significant losses in response to the FTX crisis, the traditional financial markets, as represented by the S\&P 500, exhibited contrasting behavior:
\begin{itemize}
    \item On November 11, 2022, while cryptocurrencies faced sharp declines, the \textbf{S\&P 500 (GSPC)} experienced a rise of \textbf{1.00\%}, demonstrating a decoupling from cryptocurrency market dynamics.
    \item Over the entire month of November, the S\&P 500 gained \textbf{5.81\%}, further highlighting the resilience and differing risk profiles of traditional equity markets compared to the high-risk cryptocurrency sector.
\end{itemize}

\subsubsection{FTX Conclusion}
The FTX bankruptcy served as a critical stress test, revealing the inherent volatility and risk exposure of speculative cryptocurrencies compared to the stability offered by stablecoins like Tether and traditional financial indices like the S\&P 500. This event underscores the need for robust risk management strategies and diversified investment approaches to navigate the complexities of cryptocurrency investments effectively.

\subsection{Crypto Market Analysis Synthesis}

Part I of this project delved into a comprehensive analysis of the cryptocurrency ecosystem, with an emphasis on five key cryptocurrencies (Bitcoin, Ethereum, Ripple, Dogecoin, and Tether)  and comparisons against the S\&P 500 index. Our study covered daily price behavior, volatility, correlations, and market responses to major events like the  FTX bankruptcy. Let's synthesize our key findings:

\subsubsection{Volatility and Stability}

\begin{itemize}
    \item Cryptocurrencies demonstrate substantially higher volatility than traditional markets like the S\&P 500. Dogecoin exhibited the highest volatility due to its smaller size and speculative nature.
    \item Tether's negligible volatility confirms its role as a stablecoin, offering refuge within the cryptocurrency market. 
\end{itemize} 

\subsubsection{Market Correlations}

\begin{itemize}
    \item Bitcoin, Ethereum, and other major cryptocurrencies are highly correlated, driven by similar market forces.
    \item Cryptocurrencies show low-to-moderate correlation with the S\&P 500, suggesting some independence and potential diversification benefits.
\end{itemize}

\subsubsection{FTX Crisis Impact}

\begin{itemize}
    \item The FTX bankruptcy severely impacted cryptocurrency prices, while the S\&P 500 remained largely unaffected,  highlighting sector-specific risks within crypto.
    \item November 2022's broader market picture reinforced this divergence. Cryptocurrencies declined significantly (excluding Tether), while the S\&P 500 grew, emphasizing a decoupling during cryptocurrency-specific crises. 
\end{itemize}

\subsubsection{Long-Run Market Behavior}
\begin{itemize}
    \item 2022 data illustrates that while offering potential for growth, cryptocurrencies also carry substantial risks of sharp declines.
    \item The S\&P 500's lower volatility and positive November performance underscore the  importance of traditional equity investments for risk mitigation in diversified portfolios.
\end{itemize}

\subsection{Part I Key Takeaways}

\paragraph{Diversification:} Cryptocurrencies offer diversification potential,  but investors must carefully manage their high-risk profile. 

\paragraph{Investment Strategy:} Balancing crypto holdings with safer assets like the S\&P 500 can mitigate losses during downturns.

\paragraph{Regulatory and Market Sensitivity:} Staying informed about regulatory developments and sector-specific events is crucial for navigating the dynamic cryptocurrency market. 

\subsection{Future Implications}

These insights are vital for developing robust investment strategies maximizing the potential of cryptocurrencies while safeguarding against their risks. Monitoring evolving correlations between cryptocurrencies and traditional markets will aid in understanding market dynamics and adapting investment strategies accordingly.

\hrulefill

\newpage

\section{Part II: Real-Estate Business Protocol Proposal}

\subsection{Business Proposal}

\subsubsection{Overview of the Blockchain Protocol}
Our business proposal introduces an innovative blockchain protocol designed to revolutionize the real estate sector. This protocol allows homeowners to store the deeds of their houses on the blockchain and facilitates the sale of properties without traditional intermediaries such as lawyers, brokers, or other third parties. This system not only simplifies transactions but also enhances security, reduces costs, and increases transparency.

\subsubsection{Transactional Process}

The transactional process under this protocol is streamlined to ensure efficiency and security:

\begin{enumerate}
    \item \textbf{Initiation of Sale:} Homeowners list their properties on the blockchain platform directly, bypassing the need for intermediaries. This step significantly reduces the complexity and duration of property transactions.
    \item \textbf{Proof of Ownership:} The blockchain technology inherently provides a clear, immutable record of ownership. This proof is publicly accessible and verifiable, ensuring that the current owner has indisputable ownership of the property before proceeding with the sale.
    \item \textbf{Payment and Transfer:} The buyer pays the seller in cryptocurrency, such as Bitcoin. Following payment confirmation, the property deed is automatically transferred to the buyer’s blockchain address via a smart contract, which also handles the transaction fee, typically associated with platforms like Ethereum.
    \item \textbf{Final Ownership:} The new owner receives the property deed securely stored on the blockchain, ensuring both safety and accessibility. This digital deed is resistant to tampering, loss, or theft, providing a permanent record of ownership.
\end{enumerate}

\subsubsection{Advantages of Blockchain in Real Estate}
The integration of blockchain into real estate transactions offers several improvements over traditional methods:

\begin{description}
    \item[Transparency:] The blockchain’s immutable ledger ensures that all transactions, including historical ownership data and property details (e.g., square footage, number of bedrooms, date of last renovation), are permanently recorded and openly verifiable. This level of transparency significantly reduces the potential for fraud and disputes.
    \item[Cost Efficiency:] By eliminating the need for various intermediaries and reducing paperwork and manual verification processes, the blockchain protocol cuts down on significant transactional costs. These savings make real estate transactions more economical for both buyers and sellers.
    \item[Global Accessibility:] The blockchain protocol enables international transactions without the complexities of cross-border legalities and financial transactions, opening up the property market to global participants and investors.
    \item[Market Liquidity:] The use of blockchain can enhance market liquidity. Buyers who may not have immediate access to traditional financing options can leverage decentralized finance (DeFi) solutions, such as \textbf{Aave}, for quicker funding solutions, thereby accelerating the buying process.
\end{description}

\subsection{Target Customers}

\subsubsection{Market Strategy and Consumer Segmentation}

Our blockchain protocol adopts a dual-sided market strategy designed to address distinct needs within the real estate transaction process. This approach targets two main customer segments: property sellers (and deed holders) and property buyers, along with a third, indirect segment involving financial lenders.

\paragraph{Property Sellers and Deed Holders:} The primary market segment consists of current property owners who stand to benefit substantially from blockchain integration. Traditional methods of deed storage involve physical documentation, which not only increases the risk of loss and damage but also complicates the verification and transfer processes. Our protocol offers a secure and immutable storage solution on the blockchain, eliminating the need for physical safekeeping.  

\begin{figure}[ht]
\centering
\includegraphics[width=0.8\linewidth]{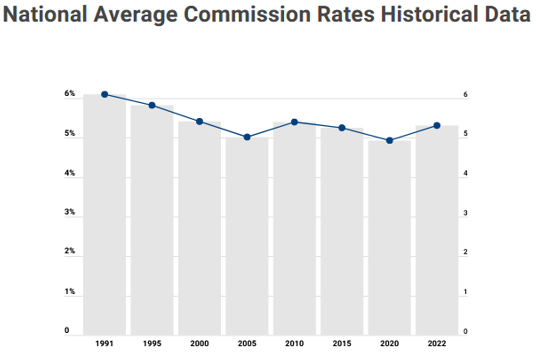}
\caption{Historical Analysis of Average Commission Rates in Real Estate Transactions}
\label{fig:avg-commission-rates}
\end{figure}

\begin{figure}[ht]
\centering
\includegraphics[width=0.8\linewidth]{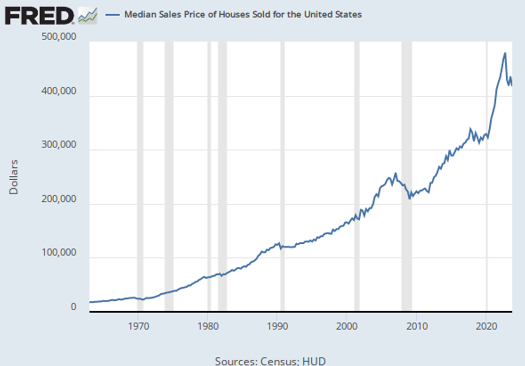}
\caption{Federal Reserve Economic Data (FRED) on Median House Sale Prices}
\label{fig:fred-median-house-sale-price}
\end{figure}

In tandem with disproportionately high the current real estate market structure necessitates multiple intermediaries, including real estate agents, brokers, and legal advisors, each adding significant transaction costs in the form of commissions and fees. Historical data indicates that commission rates have remained relatively stable over the past three decades, despite substantial increases in property values, leading to disproportionately high costs for sellers (\textit{refer to Figure \ref{fig:avg-commission-rates}}, \cite{commratetrends2013}). Moreover, median housing prices have soared, as vividly depicted in the provided data from the Federal Reserve Economic Data (FRED, \cite{fredmspus2024}) (see Figure \ref{fig:fred-median-house-sale-price}).

Our protocol simplifies this process, allowing sellers to initiate and complete sales directly on the blockchain, thereby reducing or eliminating traditional commission fees.

\paragraph{Property Buyers:}
The second primary target segment includes potential property buyers who benefit from the streamlined purchase process. Through our protocol, buyers can directly engage with sellers, conduct swift and secure transactions, and gain immediate access to verified property deeds, significantly speeding up the acquisition process. The use of smart contracts ensures that all conditions of the sale are met before the transaction is finalized, offering additional security and efficiency.

\begin{figure}[ht!]
\centering
\includegraphics[width=0.8\linewidth]{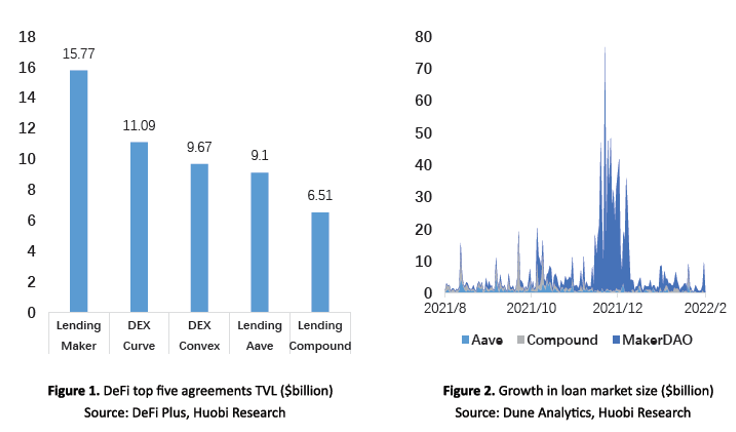}
\caption{Top Total Value Locked (TVL) in DeFi; Growth in DeFi Loan Market Size}
\label{fig:bc-lending-size}
\end{figure}

\paragraph{Financial Lenders:}
An emerging market segment within our protocol includes financial lenders, particularly those operating in the decentralized finance (DeFi) space. With the rise of blockchain technology, platforms like Aave have demonstrated significant demand for more dynamic lending solutions that offer higher yields compared to traditional financial products. Our protocol can connect these lenders directly with real estate buyers, providing a new avenue for secured lending at competitive interest rates, reflective of the increased risk profiles associated with cryptocurrency-based transactions (\textit{refer to Figure \ref{fig:bc-lending-size}}).

\subsubsection{Integration of Market Sides}
By effectively integrating these two sides of the market—sellers and buyers with the financial backing of lenders—our blockchain protocol facilitates a comprehensive ecosystem that enhances liquidity, reduces transaction latency, and improves overall market efficiency. This integrated approach not only serves the immediate participants but also introduces a scalable model for future expansions in global real estate markets.

\subsection{Competitive Analysis}

\subsubsection{Propy: A Comparative Study}

Propy emerges as a significant player within the blockchain-based real estate marketplace, providing a global platform that aligns closely with the decentralized ethos of the blockchain revolution. As a competitor, Propy's operational model is built upon eliminating traditional intermediaries from the property transaction process. By leveraging smart contracts on the blockchain, Propy ensures secure and efficient property transactions.

\paragraph{Operational Model:} The core of Propy's proposition is its decentralized marketplace, which facilitates the buying and selling of properties. This innovative approach circumvents the need for brokers and agents, potentially reducing transactional friction and cost. Furthermore, Propy offers digital deeds alongside automated escrow services, thus simplifying real estate transactions and enhancing user experience.

\paragraph{Economic Structure:} Propy's economic framework incorporates the use of its native cryptocurrency, PRO, alongside a designated fee for smart contract execution, termed PGas. The integration of PRO within their platform ecosystem not only facilitates transactional activities but also extends utility to users engaging with Propy's services. The current market valuation of PRO stands at \textbf{\$2.99}, which plays a pivotal role in the cost structure of property sales on Propy's platform.

\begin{figure}[ht!]
\centering
\includegraphics[width=0.8\linewidth]{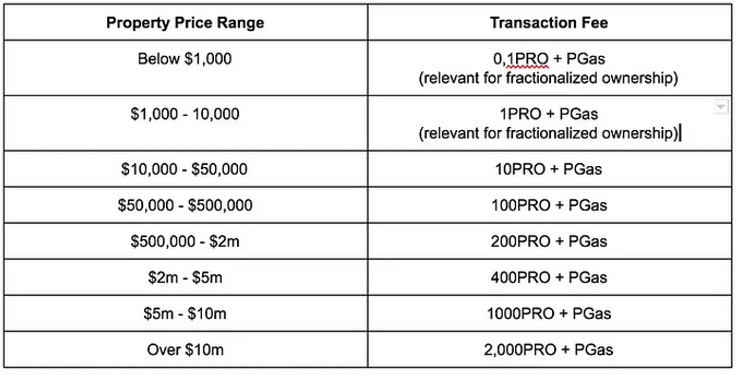}
\caption{Analysis of Propy's Transactional Fees for Property Sales}
\label{fig:propy-comp-fee}
\end{figure}

\

This competitive analysis provides a deeper understanding of Propy's strategic positioning within the blockchain-based real estate sector. By dissecting their transactional fee structure and operational model, we can assess the potential impact on our blockchain protocol's market penetration and user adoption.

\subsection{Competitive Advantage Analysis}

Our protocol presents a unique value proposition in the blockchain real estate marketplace, offering comprehensive solutions that address various stages of the real estate transaction process. It capitalizes on the inherent advantages of blockchain technology to deliver an end-to-end service that simplifies the complexities traditionally associated with real estate transactions.

\paragraph{Comprehensive Transactional Solutions:}
At the heart of our protocol is the capability to facilitate complete real estate transactions on the blockchain. This ranges from secure deed storage to the actual execution of property sales via cryptocurrency. By providing a single, unified platform, we significantly reduce the dependency on multiple services, thereby streamlining the transaction process for all stakeholders involved.

\paragraph{Transparency and Security:}
The blockchain's immutable ledger is a cornerstone feature that enhances our protocol's appeal. It serves as an unalterable record of transactions, ensuring complete transparency and security for the transaction history. This transparency is a critical factor for buyers and sellers who prioritize trust and verifiable transparency in their transactions, eliminating the traditional concerns of fraud and ambiguity in property ownership and history.

\paragraph{Cost Efficiency:}
By obviating the need for intermediaries such as agents, brokers, and legal consultants, our protocol minimizes the associated transaction costs. The conventional commission-based model, which significantly increases transaction expenses, is replaced by a more cost-effective structure that aligns with the economic preferences of a market leaning towards efficiency and reduced overhead.

\paragraph{Global Market Accessibility:}
Our protocol removes the barriers to entry for international buyers and sellers, thereby facilitating global transactions. Without the constraints imposed by legal and regulatory compliance typical of centralized systems, our platform paves the way for a more inclusive and expansive real estate market, appealing to a broader investor base and contributing to the diversity of real estate offerings.

\

In summary, our competitive advantage stems from a holistic approach that not only provides practical transactional capabilities but also fosters trust, reduces costs, and embraces global inclusivity. This strategic positioning is poised to disrupt the traditional real estate market, leveraging blockchain technology to its fullest potential.

\subsection{Implementation Strategy}

The implementation of our business model onto the blockchain comprises a systematic approach, focused on smart contract formulation, asset tokenization, oracle integration, privacy considerations, and user interface development.

\paragraph{Defining Smart Contracts:}
The foundation of our blockchain protocol is the design of smart contracts, which are digital representations of real estate assets, mortgages, and deed transfers. The contracts will encapsulate the logic for buying, selling, transferring ownership, and managing mortgage payments. This will involve:
\begin{enumerate}
    \item Structuring smart contracts to encapsulate real estate transaction requirements.
    \item Integrating functions for various transaction processes within these contracts.
\end{enumerate}

\paragraph{Tokenizing Real Estate Assets:}
We will transform real estate properties into digital tokens on the Ethereum blockchain, where each token signifies property ownership. This process will:
\begin{enumerate}
    \item Utilize established token standards, such as ERC-20, for the representation of real assets in the digital domain.
    \item Deploy contracts to issue and regulate these tokens, assigning unique identifiers to each property (\cite{erc20whitepaper}).
\end{enumerate}

\paragraph{Integration of Oracles:}
Oracles will be employed to incorporate off-chain data, like property specifications and legal documents, into the blockchain. These oracles will:
\begin{enumerate}
    \item Source trusted data essential for executing real estate transactions.
    \item Update on-chain records to reflect accurate off-chain information, ensuring the veracity and reliability of data (\cite{blockchainoracle2020}).
\end{enumerate}

\paragraph{Privacy Enhancements:}
Our protocol will incorporate privacy measures to safeguard sensitive transaction data, allowing for public verification while maintaining confidentiality. We will:
\begin{enumerate}
    \item Develop smart contracts with robust privacy controls.
    \item Implement encryption techniques to restrict data access to authorized entities only.
\end{enumerate}

\paragraph{User Interface Development:}
To facilitate user interaction with our blockchain platform, we will develop accessible and intuitive interfaces. These interfaces will:
\begin{enumerate}
    \item Offer a seamless user experience for property searches, transaction initiation, mortgage tracking, and deed management.
    \item Provide tools that are comprehensible and efficient for users, irrespective of their familiarity with blockchain technology.
\end{enumerate}

\

The strategic deployment of these elements will result in a robust blockchain protocol for real estate, streamlining the transaction process and enhancing the overall experience for users in the real estate market. The careful orchestration of smart contracts, tokenization, oracles, privacy considerations, and user interfaces are essential components of our strategy to integrate real estate transactions with blockchain technology effectively.

\subsection{Economic Viability for Ethereum}

Evaluating the economic feasibility of our business proposal on the Ethereum platform involves careful consideration of various factors, including scalability, transaction fees, and the complexity of smart contracts.

\paragraph{Scalability Concerns:}
Ethereum's scalability is a pivotal concern, particularly as transaction volumes escalate. As the leading blockchain platform, Ethereum faces challenges in maintaining performance amid rising demand. The introduction of Ethereum 2.0 promises to alleviate these issues through sharding and a proof-of-stake consensus mechanism, potentially enhancing throughput and lowering transaction costs (\cite{blockchaingasfees2021}).

\paragraph{Transaction Fees:}
Gas fees on Ethereum are known for their volatility and can constitute a significant portion of transaction costs. These fees tend to surge during periods of network congestion, impacting the cost-benefit analysis for users engaging in real estate transactions. Monitoring the historical trends of Ethereum's average gas fee is crucial in forecasting and managing the financial viability of transactions (\cite{ethereumgasprices2024}).

\begin{figure}[ht]
\centering
\includegraphics[width=0.8\linewidth]{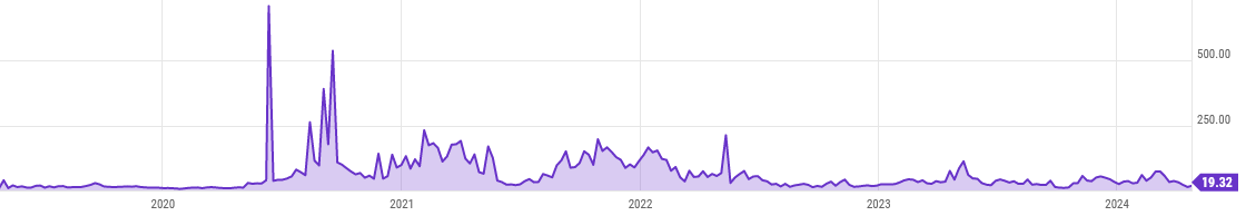}
\caption{Average Ethereum Gas Fees Over the Last Five Years}
\label{fig:avg-eth-gas-fees}
\end{figure}

\paragraph{Smart Contract Complexity:}
The smart contracts at the core of our real estate protocol, which will enable property transfers, loan repayments, and the enforcement of covenants, are inherently complex. The intricacy of these contracts is directly proportional to the computational resources required, thus influencing the overall gas fees incurred. This complexity must be carefully managed to ensure that the benefits of using the Ethereum platform outweigh the costs for all parties involved.

\

The prospective enhancements with Ethereum 2.0 alongside strategic management of smart contract complexity and monitoring of gas fees are essential for ensuring the economic viability of deploying our real estate transaction protocol on the Ethereum blockchain. As we progress, it will be imperative to remain adaptable to the evolving blockchain landscape to maintain a competitive and cost-effective platform for real estate transactions.

\subsection{Comparative Analysis of Blockchain Platforms}

\begin{figure}[ht]
\centering
\includegraphics[width=0.8\linewidth]{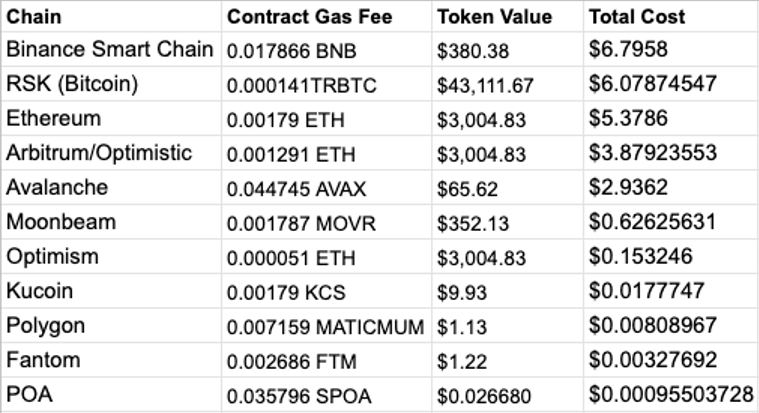}
\caption{Comparative Analysis of Ethereum Versus Alternative Blockchain Platforms}
\label{fig:eth-vs-misc}
\end{figure}

In exploring the economic viability and technical suitability of our real estate transaction protocol, we extend our analysis beyond Ethereum to other leading blockchain platforms, each with distinct attributes and potential advantages.

\paragraph{Binance Smart Chain:}
Binance Smart Chain (BSC) emerges as a viable alternative, promising higher throughput and lower latency compared to Ethereum, which is advantageous for high-demand scenarios. Despite these benefits, BSC's more centralized nature may raise concerns among stakeholders seeking a fully decentralized solution.

\paragraph{Solana:}
Solana presents a compelling case for applications necessitating rapid transaction processing, offering superior transaction speeds and scalability (\cite{solana2022}). While Solana provides an efficient alternative to Ethereum, its developer ecosystem and tooling are less mature, potentially imposing limitations on development and integration efforts.

\paragraph{Polkadot:}
The multi-chain framework of Polkadot facilitates cross-chain interoperability, which can significantly enhance the scope and flexibility of our protocol. Polkadot's design allows for seamless integration with a variety of blockchain networks, potentially expanding the protocol's reach. Nevertheless, Polkadot's infrastructure and tooling are still evolving, which may introduce challenges during early adoption phases.

\

This comparative analysis underscores the importance of selecting a blockchain platform that aligns with our protocol's requirements for security, decentralization, transaction speed, and scalability. As we proceed with our implementation strategy, ongoing evaluation of these platforms' evolving capabilities will be imperative to maintain an innovative and user-centric service in the dynamic real estate market.

\subsection{Synthesis of the Blockchain Real-Estate Protocol}

The business protocol presented in Part II encapsulates an innovative, blockchain-based approach to real estate transactions. The proposed system introduces a transformative model that empowers homeowners to engage directly in the sale and purchase of properties, effectively circumventing the traditional, intermediary-reliant processes that are often cumbersome and less secure. 

\paragraph{Overview and Efficacy of Protocol Application}
The cornerstone of the proposed protocol lies in its utilization of blockchain's inherent properties such as immutability, transparency, and distributed consensus. These properties facilitate a seamless transition of deeds and payments, providing a robust proof of ownership and streamlining transactions. By leveraging smart contract technology, the protocol ensures that all prerequisites of a property transaction are automatically met, heralding a new era of efficiency in property dealings.

\paragraph{Enhancing Real Estate Transaction Dynamics}
The protocol offers multiple benefits over conventional methods. It provides a public, transparent ledger for ownership and transaction history, reduces the costs associated with property transactions by eliminating intermediary fees, and enables global participation in the real estate market. Additionally, the protocol introduces greater liquidity to the market by integrating with decentralized financial platforms, allowing prospective buyers to secure funding rapidly.

\paragraph{Consumer-Centric Market Strategy}
This protocol advocates a dual-sided market strategy that aims to reform the real estate transaction paradigm. Property sellers are afforded the ability to secure deed storage on the blockchain while benefiting from direct market access for sales, effectively bypassing intermediary overheads. For property buyers, the protocol simplifies the purchase process, offering immediate access to property details and streamlining the transfer of ownership. Moreover, financial lenders find a new marketplace in which to offer secured loans, augmented by blockchain's security features.

\paragraph{Competitive Landscape and Advantages}
In the competitive landscape, the proposed protocol differentiates itself by presenting a comprehensive and integrated solution that extends beyond mere transaction facilitation. It anticipates the current and future needs of the real estate market, focusing on user experience, transactional integrity, and market inclusivity. Against competitors like Propy, RealT, and Deedcoin, the protocol asserts its edge through its amalgamation of transactional efficiency, cost savings, and market reach.

\paragraph{Strategic Implementation and Economic Considerations}
The practical implementation of the protocol on the Ethereum blockchain, and the considerations for its economic viability, are outlined with a foresight into potential scalability issues and transaction costs. Alternative blockchains such as Binance Smart Chain, Solana, and Polkadot are appraised for their suitability, with an emphasis on their comparative advantages in terms of transaction speed, costs, and infrastructural development. 

\paragraph{Implications and Future Prospects}
In conclusion, the blockchain real-estate protocol promises to not only revolutionize the manner in which real estate transactions are conducted but also to serve as a blueprint for future applications of blockchain technology in other domains. The synthesis of the protocol's operational model, strategic implementation, and competitive positioning underscores its potential to offer a superior alternative to the established real estate transaction processes, setting a new benchmark in efficiency, security, and global accessibility.

\section{Conclusion}

This paper has presented an in-depth examination of the cryptocurrency market, followed by a pioneering proposal for a blockchain-based real estate transaction protocol. Through meticulous analysis, it has provided evidence of the intricate dynamics governing cryptocurrency price movements, volatility, and correlations, particularly in the context of the FTX bankruptcy event, thereby illuminating the vulnerabilities and resilience inherent in the digital currency landscape. In tandem, it has offered a visionary blueprint for the utilization of blockchain technology in streamlining real estate transactions, proposing a model that is poised to redefine the sector.

\subsection{Synthesis of Findings}
The market analysis revealed that cryptocurrencies are not only highly volatile but are also subject to correlated movements, which can lead to systemic risks within the digital asset class. Yet, this volatility and interconnectivity also underscore the potential of cryptocurrencies to diversify investment portfolios when judiciously balanced with traditional assets. The FTX bankruptcy served as a litmus test for the market, delineating the stability offered by stablecoins and the S\&P 500 in contrast to the pronounced volatility of cryptocurrencies like Bitcoin and Dogecoin. 

On the frontier of innovation, the blockchain real estate proposal detailed in Part II of this paper is set to disrupt a long-established industry. By removing intermediaries, reducing transaction costs, and enhancing transparency, the protocol demonstrates a tangible application of blockchain beyond speculative trading, embodying the technology’s transformative potential in real-world asset management and exchange.

\subsection{Prospects and Implications}
The synthesis of the paper's findings paints a nuanced picture of the cryptocurrency market’s complexities and introduces a sophisticated approach to real estate transactions that capitalizes on blockchain technology's strengths. As the cryptocurrency market continues to mature, it is anticipated that investor strategies will adapt to encompass both traditional and digital assets, ensuring balanced portfolios that mitigate risk while capitalizing on growth opportunities.

The real estate blockchain protocol, while nascent, holds promise for a radical shift in property ownership transfer, marking a significant leap towards a more interconnected, efficient, and accessible global market. Its implications extend beyond the real estate sector, signaling the advent of a broader adoption of blockchain in various facets of commerce and governance, ushering in a new era of decentralized digital solutions.

\subsection{Summary and Forward Outlook}
In summary, the research presented in this paper contributes meaningfully to the understanding of cryptocurrencies and offers a progressive application of blockchain technology. As we witness the convergence of traditional financial methodologies with groundbreaking digital solutions, the potential for innovation in both markets and technology is boundless. Future research and development will undoubtedly continue to expand on these foundations, further integrating the burgeoning possibilities of blockchain technology into the fabric of societal and economic structures.

In moving forward, continuous monitoring of market trends, regulatory developments, and technological advancements will be crucial in optimizing the strategies and applications discussed herein. The cryptocurrency market's evolution and the blockchain real estate protocol's maturation will undoubtedly serve as critical barometers for the future trajectory of digital finance and property transactions. The journey ahead promises to be as challenging as it is exciting, with the potential to redefine the very essence of investment, ownership, and exchange in an increasingly digital world.

\printbibliography

\end{document}